\documentclass[fleqn,usenatbib]{mnras}

\usepackage{newtxtext,newtxmath}

\usepackage[T1]{fontenc}

\DeclareRobustCommand{\VAN}[3]{#2}
\let\VANthebibliography\thebibliography
\def\thebibliography{\DeclareRobustCommand{\VAN}[3]{##3}\VANthebibliography}

\usepackage{graphicx}	
\usepackage{amsmath}	
\usepackage{makecell}
\usepackage{multirow}  
\usepackage{tabularx}  

\newcommand{\wcen}{$\omega$ Cen}

\usepackage{eso-pic}
\AddToShipoutPictureBG*{%
  \AtPageUpperLeft{%
    \hspace{0.75\paperwidth}%
    \raisebox{-4.5\baselineskip}{%
      \makebox[0pt][l]{\textnormal{FERMILAB-PUB-22-890-T}}
}}}%

\title[On the gamma-ray emission of Sgr]{On the gamma-ray emission from the core of the Sagittarius dwarf galaxy}

\author[Addy J. Evans et al.]{
 Addy J. Evans$^{1}$, Louis E. Strigari$^{1}$, Oskar Svenborn$^{2}$, Andrea Albert$^{3}$, J. Patrick Harding$^{3}$,
 \newauthor{ Dan Hooper$^{4, 5}$, Tim Linden$^{2}$}, Andrew B. Pace$^{6}$
\\
$^{1}$Mitchell Institute for Fundamental Physics and Astronomy, Department of Physics and Astronomy, Texas A\&M University, College Station, TX, 77845\\
$^{2}$Stockholm University and The Oskar Klein Centre for Cosmoparticle Physics, Alba Nova, 10691 Stockholm, Sweden\\
$^{3}$Physics Division, Los Alamos National Laboratory, Los Alamos, NM, USA\\
$^{4}$Theoretical Astrophysics Group, Fermi National Accelerator Laboratory, Batavia, IL, 60510, USA\\
$^{5}$Department of Astronomy \& Astrophysics and the Kavli Institute for
Cosmological Physics (KICP), University of Chicago, Chicago, IL, 60637, USA\\
$^{6}$McWilliams Center for Cosmology, Carnegie Mellon University, 5000 Forbes Ave, Pittsburgh, PA 15213, USA\\
}

\pubyear{2015}

\begin{document}
\label{firstpage}
\pagerange{\pageref{firstpage}--\pageref{lastpage}}

\maketitle


\begin{abstract}
We use data from the Large Area Telescope onboard the Fermi gamma-ray space telescope (Fermi-LAT) to analyze the faint gamma-ray source located at the center of the Sagittarius (Sgr) dwarf spheroidal galaxy. In the 4FGL-DR3 catalog, this source is associated with the globular cluster, M54, which is coincident with the dynamical center of this dwarf galaxy. We investigate the spectral energy distribution and spatial extension of this source, with the goal of testing two hypotheses: (1) the emission is due to millisecond pulsars within M54, or (2) the emission is due to annihilating dark matter from the Sgr halo. For the pulsar interpretation, we consider a two-component model which describes both the lower-energy magnetospheric emission and possible high-energy emission arising from inverse Compton scattering. We find that this source has a point-like morphology at low energies, consistent with magnetospheric emission, and find no evidence for a higher-energy component. For the dark matter interpretation, we find that this signal favors a dark matter mass of $m_{\chi} = 29.6 \pm 5.8$ GeV and an annihilation cross section of $\sigma v = (2.1 \pm 0.59) \times 10^{-26} \,\text{cm}^3/$s for the $b \bar{b}$ channel (or $m_{\chi} = 8.3 \pm 3.8$ GeV and $\sigma v =  (0.90 \pm 0.25) \times 10^{-26} \, \text{cm}^3/$s for the $\tau^+ \tau^-$ channel), when adopting a J-factor of $J=10^{19.6} \, \text{GeV}^2 \, \text{cm}^{-5}$. This parameter space is consistent with gamma-ray constraints from other dwarf galaxies and with dark matter interpretations of the Galactic Center Gamma-Ray Excess.

\end{abstract}




\section{Introduction} 
\par The Sagittarius (Sgr) dwarf spheroidal (dSph) galaxy is one of the closest and most luminous satellite galaxies orbiting the Milky Way (MW). This dSph has both a discernible core, as well as a long tidal tail which spans the entirety of the sky, extending more than 100 kpc~\citep{2003Majewski,2004ASPC..327..239L}.
The half-light radius of the core, $\sim$1.5 kpc, is among the largest of all dSphs. The kinematics of the stars within the core region, in combination with models that aim to match the properties of the tidal tails, provide strong evidence that the central region of Sgr is dominated by dark matter (DM)~\citep{Ibata1997AJ....113..634I,2010ApJ...725.1516L,2011ApJ...727L...2P}.

\par Sgr is unique amongst MW dSphs (with the exception of Fornax) in that it has an associated population of globular clusters (GCs). The most prominent GC associated with Sgr is M54, which coincides with the center of the Sgr core. Several other GCs have long been associated with Sgr, including Arp 2, Terzan 7, Terzan 8, Palomar 12, Whiting 1, NGC 2419, and NGC 5824~\citep{2019A&A...630L...4M,2020MNRAS.498.2472K}. In addition to these, there is recent evidence from the Via Lactea Extended Survey (VVVX) near-infrared database for an additional population of GCs associated with the core of Sgr~\citep{Minniti2021_1,Minniti2021_2}. Up to 20 new GC candidates have been identified in VVVX, several of which are considered to be high probability candidates due to their measured overdensities of RR Lyrae stars. Including these new discoveries, Sgr is now the dSph with the largest number of associated GCs.  

\par Multiwavelength observations can provide us with a more detailed understanding of the GC population and DM halo of Sgr.
There have been several studies of the Sgr/M54 region in the X-ray regime which suggest that cataclysmic variable stars and low-mass X-ray binaries are each present within M54~\citep{2006Ramsay_M54}.
Although there have been similar searches within the dwarf's main body~\citep{2006Ramsay_xraySgr}, the number of X-ray sources observed in that region has been consistent with the expected number of background sources. 
Gamma-ray studies have also been conducted in the Sgr region, although typically as one of several stacked sources in searches for DM annihilation products~\citep{2012ApJ...746...77V,2014PhRvD..90k2012A,2014PhRvD..89d2001A,2015JCAP...09..016H}.
Since Sgr has no detected HI gas associated with its central core~\citep{Grcevich:2009gt}, the only sources of $\gtrsim 100$ MeV gamma-ray emission (other than DM) would be millisecond pulsars (MSPs).
In this way, Sgr is unique, as it possesses a DM halo that could produce gamma-ray photons from DM self-annihilation as well as a population of GCs, which are often gamma-ray bright due to their MSP populations~\citep{Fermi2010}.

\par Field MSPs (e.g. those not associated with GCs and in the main body of the dSph) could also produce detectable fluxes of gamma-rays.~\citet{2016Winter} used the stellar masses of classical dSphs to estimate the gamma-ray luminosity functions of their field MSP populations. While Sgr was not included in that study, its stellar mass of $\sim 4 \times 10^8$ M$_\odot$ is most similar to that of Fornax's, $\sim 2 \times 10^7$ M$_\odot$, which the authors find to be just below the threshold for detection. This suggests that it may be possible to detect the gamma-ray emission from Sgr's MSP population with current Fermi-LAT data. 
The authors of that study also compare this prediction to the gamma-ray flux expected from the annihilations of a 30 GeV DM particle (to $b\bar{b}$) with a cross section of $3 \times 1 0^{-26}$ cm$^{3}$/s. In this comparison, the authors found that the two predicted fluxes are nearly indistinguishable in the case of Fornax. Given its stellar mass, distance, and DM content, this result implies that Sgr could be visible due to MSP emission, DM annihilation, or both. 

\par While the spectral energy distribution (SED) of the gamma rays observed from MSPs is similar to that predicted from the annihilation of $\sim$ 20-50 GeV DM particles~\citep{Baltz2007,2013MNRAS.436.2461M}, one can attempt to differentiate between these potential signals by considering their different morphologies.
Given the radius ($\sim 50 \, {\rm pc}$) and distance ($\sim$ 26.5 kpc) of M54~\citep{2009AJ....137.4478K,2021MNRAS.505.5957B,Ferguson2020}, any gamma-ray emission from this GC would likely be indistinguishable from a point source to Fermi-LAT. In contrast, any gamma-ray emission from DM annihilating in Sgr's halo would be more spatially extended, potentially at a level that could be detected by Fermi, depending on the details of the DM distribution.

\par The fact that Sgr is one of the nearest dSphs makes it a promising target for DM searches using gamma rays.  However, because Sgr is located just below the Galactic center and in a region with significant Galactic diffuse gamma-ray emission, it has been the subject of relatively few studies searching for the products of DM annihilation, at least compared to other dSphs (see, however,~\citet{2012ApJ...746...77V,2014PhRvD..90k2012A}). 
In addition, the likely non-equilibrium nature of Sgr's dynamical state makes it more difficult to interpret its stellar kinematics and extract a reliable determination of its DM distribution. Nonetheless, now that Fermi-LAT has accumulated over 13 years of data from this region, it is prudent to reconsider Sgr as a possible gamma-ray source.

MSPs (and any GCs containing MSPs) produce two distinct components of gamma-ray emission. The first of these is the radiation that is produced by charged particles traveling along the open magnetic field lines of
a pulsar. This prompt or ``magnetospheric'' emission peaks at $\sim$ GeV energies with a characteristic log-parabola shape. The second component, which dominates at high energies, is thought to arise from the inverse Compton scattering (ICS) of very high-energy electrons/positrons which escape into the surrounding environment. Observations by the High Altitude Water Cherenkov (HAWC) Observatory and the Large High Altitude Air Shower Observatory (LHAASO) have shown that young and middle-aged pulsars are typically surrounded by bright, spatially-extended, multi-TeV emitting regions known as ``TeV halos''~\citep{Hooper:2017gtd,Linden:2017vvb,HAWC:2021dtl,HAWC:2019tcx}. Even more recently, it has been shown (at the 99\% C.L.) that millisecond pulsars also generate TeV halos~\citep{Hooper:2021kyp}. Further supporting this conclusion,~\cite{Song2021} have recently conducted an analysis in which the authors detected, at 8.2 $\sigma$,
a high-energy ($> 10$ GeV) power-law component of gamma-ray emission in the spectra of gamma-ray bright
globular clusters.
These results are most
naturally interpreted as evidence for an ICS component in addition to the magnetospheric
gamma-ray emission from GCs. The ratio of the observed luminosities of these two components can vary significantly among GC's, in cases being as small as $L_{\text{IC}} / L_{\gamma} \leq 0.07$ or as large as $L_{\text{IC}} / L_{\gamma} \geq 6.40$, reflecting variations associated with the beaming of the magnetospheric emission, or potentially arising from additional environmental factors or unaccounted for emission mechanisms (such as synchrotron or bremsstrahlung).

\par The Fermi Collaboration's most recent source catalog (4FGL-DR3) contains a gamma-ray source that is coincident with the spatial location of M54~(\citealt{20224FGL3DR}; and confirmed by~\citealt{2022RAA....22k5013Y}). In addition,~\citet{Crocker2022}~have reported evidence for gamma-ray emission that is approximately 4 degrees offset from the main body of this dSph. The authors of that study further describe this emission to be extended, approximately $\sim 20^\circ $ in diameter, and highly statistically significant, $\ge 5 \sigma$. 
The reported SED of this source has an intensity at $\sim 1 \,{\rm GeV}$ that is comparable to that observed at $\sim 100 \, {\rm GeV}$ (in GeV cm$^{-2}$ s$^{-1}$ sr$^{-1}$ units). 
The authors interpret this emission as originating from MSPs, adopting a model that includes both magnetospheric emission and high-energy emission arising from ICS.

\par In this paper, we analyze the region of the sky centered on the Sagittarius dwarf galaxy using Fermi-LAT gamma-ray data. To characterize the gamma-ray emission from within our region of interest (ROI), we test for both point-like and extended emission from the Sgr/M54 region itself and search for unassociated sources that belong to the Sgr system. 
We confirm the existence of the point source associated with M54, as first identified by the Fermi-LAT Collaboration, and subsequently search for high-energy ($>10$ GeV) emission associated with this source.

We then test a DM annihilation scenario and derive constraints on the DM's annihilation cross section and mass.

\par The remainder of this paper is organized as follows. 
In section~\ref{sec:data}, we outline the Fermi-LAT data and software used in our analysis and discuss the methodology of our point-like and diffuse-like tests, as well as our search for other Sgr-associated point-like sources.
In section~\ref{sec:results}, we discuss the results of our analysis, focusing first on the GC/MSP interpretation of this emission.
We subsequently discuss our results within the context of annihilating DM in section~\ref{sec:dm}.
Lastly, we summarize our conclusions in section~\ref{sec:final}.

\section{Data Analysis}
\label{sec:data} 

\par In this study, we perform both a point source and an extended source binned likelihood analysis, centered on M54, using the Fermitools 2.0.8.\footnote{https://github.com/fermi-lat/Fermitools-conda/wiki}
We utilize~\texttt{FermiPy}~\citep{Wood:2017yyb}, which is a python-based software package that automates the tools for Fermi-LAT source analysis. 
For our data selection, we use~\texttt{Pass 8 SOURCE}-class photon events with the corresponding instrument response functions,~\texttt{P8R3}\textunderscore\texttt{SOURCE}\textunderscore\texttt{V3}. 
We select both \texttt{FRONT} and \texttt{BACK} converting events (evclass=128 and evtype=3), with energies in the range 300 MeV to 500 GeV. 
For our primary analysis, we exclude photons with energies below 300 MeV in order to avoid complications associated with the broader point spread function (PSF)~\citep{20224FGL3DR}.
We use approximately 13.5 years of data, corresponding to mission elapsed times between 239557417 s and 661506150 s. 
We apply the recommended \texttt{(DATA\char`_QUAL>0)\&\&(LAT\char`_CONFIG==1)} filter to ensure quality data and a zenith cut of $z_{max} = 90^\circ$ to filter background gamma-ray contamination from the Earth's limb. 
\par We consider a $15^\circ \times 15^\circ$ ROI centered on M54. 
For our likelihood maximization, we take a 0.1$^\circ$ angular pixelation and use the \texttt{MINUIT} optimizer method within \texttt{gtlike}. 
We use an input source model that includes all sources in the 4FGL-DR3 catalog~\citep{20224FGL3DR}~out to a square of $20^\circ \times 20^\circ$. 
Including sources beyond the ROI ensures that sources on the edge of the ROI are properly modeled.
For the interstellar emission model we use the recommended \texttt{gll}\_\texttt{iem}\_\texttt{v07.fits}, and for the isotropic emission we use \texttt{iso}\_\texttt{P8R3}\_\texttt{SOURCE}\_\texttt{V3}\_\texttt{V1.txt}. 

\par In a general \texttt{FermiPy} analysis, one defines the model sources within the ROI and then performs multiple likelihood tests to determine the best-fitting parameters of the model sources. 
In this case, we define the Test Statistic (TS) as $\text{TS} = -2 \ln (\mathcal{L}_0/\mathcal{L}_1) $ where $\mathcal{L}_0$ represents the likelihood of the null hypothesis and $\mathcal{L}_1$ represents the likelihood of the alternative.
Furthermore, as is typically done, we assume that Wilks' Theorem applies such that the log likelihoods follow a normal distribution and that the statistical significance (in standard devitations) is given by $\sqrt{\text{TS}}$.

\par 
In the subsections below, we discuss two approaches to our analysis of the M54/Sgr region. 
Our first approach entails a point source analysis of the region to characterize M54 and any other possible sources of interest which could be attributed to Sgr. In our second approach, we search for evidence of extended emission from the M54/Sgr system. 

\subsection{Point source analysis}

We first perform a point source analysis of the region. Our initial model consists of the aforementioned Fermi-LAT background models as well as the 4FGL-DR3 catalog sources. We keep the spectral types of all sources fixed to their catalog values except for our source of interest.

\par Due to the location of M54 on the sky (just south of the Galactic center), the possibilty of source contamination, especially at the lower end of the Fermi-LAT energy sensitivity (see, for example,~\citet{BalletICRC2015}), is a significant complication. In particular, it is not always straightforward to reliably separate faint or extended sources within the ROI from diffuse background emission.

\par For the spectrum of the gamma-ray emission from M54, we consider several parameterizations:
\begin{enumerate}
    \item A simple power law (ICS model),
    \begin{equation}
       \frac{dN}{dE} = \text{N}_1  \left(\frac{E}{E_0} \right)^{\gamma_1},
       \label{eqn:ics}
    \end{equation}
    where $N_1$ is the normalization of the flux, $E_0$ is the scale energy, and $\gamma_1$ is the spectral index.
    \item A power law with an exponential cut-off (curvature or magnetospheric emission model, CRV),
    \begin{equation}
        \frac{dN}{dE} = \text{N}_2  \left(\frac{E}{E_0} \right)^{\gamma_2} \exp{\left[-\left(\frac{E}{E_\text{cut}} \right)\right]},
        \label{eqn:mag}
    \end{equation}
    where the energy cut-off is an additional parameter, $E_\text{cut}$, and N$_2$ is the normalization.
    \item A two-component model which is a combination of a simple power law and magnetospheric emission,
    \begin{equation}
            \frac{dN}{dE} = \text{N}_1  \left(\frac{E}{E_0} \right)^{\gamma_1}
            +
            \text{N}_2  \left(\frac{E}{E_0} \right)^{\gamma_2} \exp{\left[-\left(\frac{E}{E_\text{cut}} \right)\right]}. 
        \label{eqn:combo}
    \end{equation}
\end{enumerate}

Note that while a log-parabola parameterization is sometimes adopted for the magnetospheric emission from MSPs and GCs~\citep{20224FGL3DR}, we've chosen to adopt the form described above to more easily compare our results to those of~\citet{Song2021}.

\par We test the robustness of our two-component model by applying it to the bright GC, Terzan 5, which is known to produce high-energy emission. We have selected Terzan 5 for this test for several reasons. First, Terzan 5, like M54, is one of the most massive GCs in the MW. It is also thought to be the remnant of a nuclear star cluster at the center of a progenitor dwarf galaxy~\citep{2009Nature_Ter5}, similar to M54. Second, Terzan 5 is the only GC to be detected at very-high-energies~\citep{Ter5HESS}, featuring emission that extends to at least 10 TeV. 
Terzan 5 is also included in the study of~\citet{Song2021}, in which they detect an ICS component with a luminosity that is comparable to that of its CRV component, $L_{\text{ICS}} / L_{\text{CRV}} = 0.37$.

\par Once our model has been defined, we proceed to determine the best-fit normalization for all of the 4FGL catalog sources, the diffuse emission components, and for the emission from M54. 
We begin this process by running the \texttt{FermiPy} method \texttt{gta.optimize} on the ROI.
We then free all of the spectral parameters of the M54 source(s) and fit to them individually using the \texttt{FermiPy} wrapper of the \texttt{pyLikelihood} fitting routine, \texttt{gta.fit}, while keeping the background fixed to the best-fitting values found in the original optimization.
Finally, we run the method \texttt{gta.sed} on our M54 source(s). 
From this method, we can determine the TS of different spectral models for our source.

\par In practice, the way we implement a two-component source is by removing the 4FGL-DR3 M54 catalog source and replacing it with two sources at the same location, one with a spectral type defined by Equation~\ref{eqn:ics}~and the other with a spectral type defined by Equation~\ref{eqn:mag}. 
For the likelihood fits, we first set the spectral parameters of each source to match the best-fitting values found by~\citet{Song2021} in their universal fitting method.
For the CRV source, these values are $\gamma_2 = 0.88$ and $\log_{10}(\text{E}_\text{cut}) = 3.28$ MeV, while for the power law source, we set $\gamma_1 = 2.79$.
We then allow the normalization and spectral parameters to float for each source simultaneously to determine their contributions to the total integrated photon flux.


\begin{figure}
    \includegraphics[width=\columnwidth]{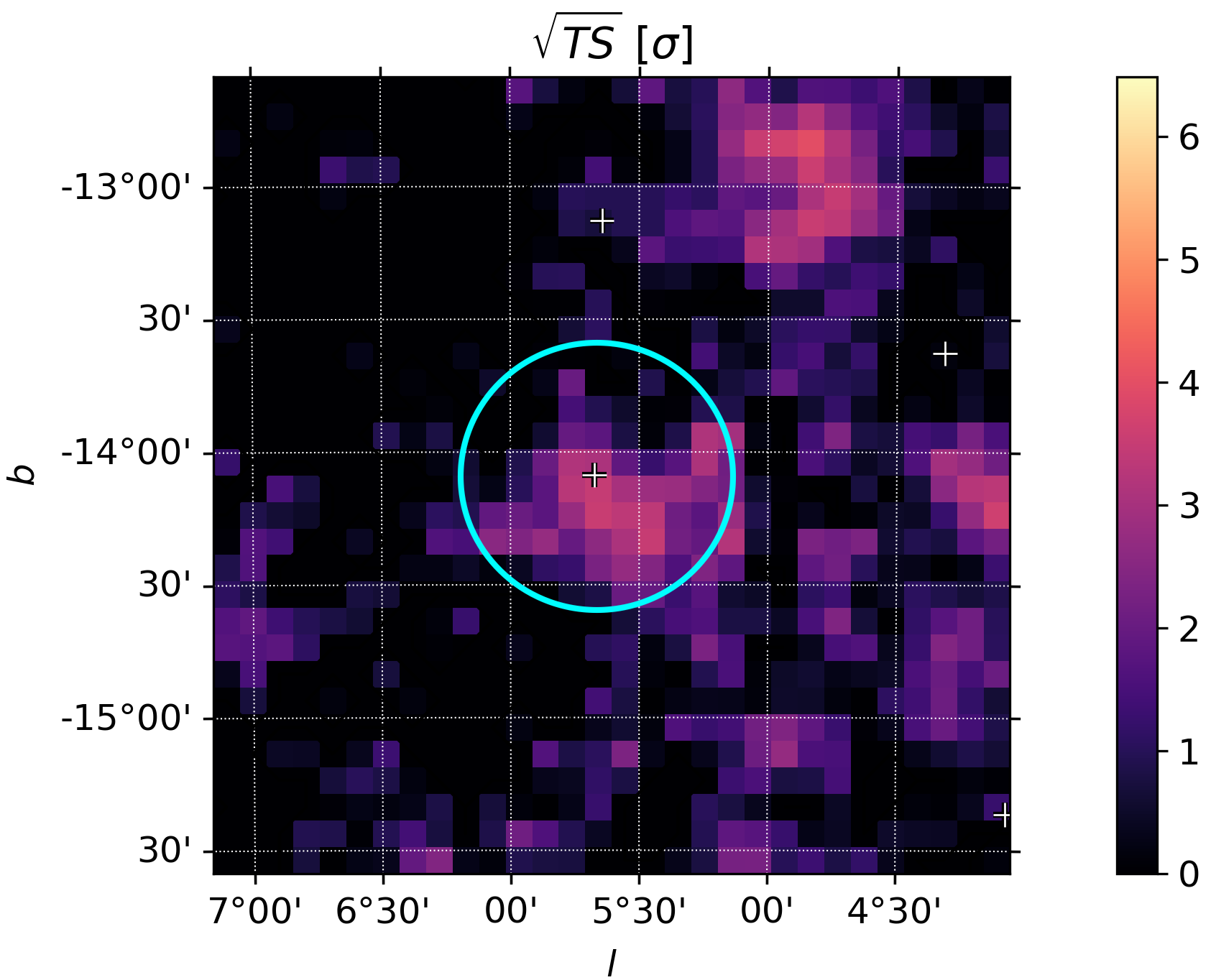}
    \caption{Test statistic (TS) map of a zoomed in region of interest centered on the Sagittarius/M54 system. The white cross in the center of the figure represents the 4FGL-DR3 catalog location of the source associated with M54, 4FGL-J1855.1-3025. Other white crosses denote sources which have been accounted for in the modeling. The cyan circle shows the half-light radius of M54. The source coincident with M54 is detected at a level of $\sqrt{\text{TS}} \sim 4.5 - 5.0$ (see text for details).}
    \label{fig:tsmap}
\end{figure}

\subsection{Unassociated source analysis}

Thus far, we have described our analysis of the single, point-like source coincident with the core of Sgr. Next, we performed a search for other sources of gamma-ray emission which could be associated with this dSph. As there are no known pulsars in Sgr or M54~\citep{1996MNRAS.282..691B}, and Sgr has no active star formation or gas~\citep{1999A&A...349....7B,1994MNRAS.270L..43K}, we compare the locations of our unassociated point sources to the locations of Sgr's GCs. 
We search for spatial coincidences by comparing the locations of Sgr's GCs~\citep{Minniti2021_1, Minniti2021_2,2010AJ....140.1830G} to both sources labeled as unassociated in the 4FGL catalog, as well as unassociated peaks in TS space within the ROI.

\par To find unassociated peaks in TS space, we use the \texttt{FermiPy} function~\texttt{gta.find\_sources}. In order to identify any possible sources near the threshold of detection, we set a low threshold of TS $\geq 9$.
Then, to better constrain the locations of the unassociated catalog sources and the newly found sources, we use the function \texttt{gta.localize}. 
The best-fitting position for the source of interest is then updated, which we compare to the locations of known GCs within Sgr.

\subsection{Extended source analysis}

\citet{Crocker2022}~report the high-significance (up to $\sim 23\sigma$) detection of an extended source with a best-fit location that is centered $\sim 4^\circ$ from M54. 
In this study, we also search for extended emission, focusing on templates that are centered at the location of the core of Sgr. To this end, we first perform a basic extension test with the built-in \texttt{FermiPy} tool, \texttt{gta.extension}, and the 4FGL catalog background models.\footnote{https://fermi.gsfc.nasa.gov/ssc/data/access/lat/BackgroundModels.html}
We test two spatial templates: one where the spatial morphology is described by a flat and uniform disk, and another where the spatial morphology is described by a 2D Gaussian.
In each case, we consider templates which are extended by up to 1$^\circ$ in radius. 
As we did in our point source analysis, we keep the background and other sources fixed.
In section~\ref{sec:dm}, we consider additional extended templates which are motivated by scenarios involving annihilating DM.

\section{Results}
\label{sec:results}
\subsection{Detection of the M54 Point Source}

\par In Figure~\ref{fig:tsmap}, we show a TS map of a region within our ROI for the energy range of [300 MeV, 500 GeV].
The point source coincident with M54 (4FGL-J1855.1-3025) is detected with a TS of 21.9 for this energy range and with TS=24.3 for [100 MeV, 500 GeV], adopting the log parabola spectral model. Note that this is the TS that is obtained after optimizing the spectrum of this source, and before performing any other fitting procedures.
While the results of our spectral analysis defined in section~\ref{sec:SED}~assume an energy range of [300 MeV, 500 GeV], we note that the 4FGL-DR3 catalog reports a TS of $\sim 26$ for the M54 source within the energy range of [100 MeV, 500 GeV].
For this energy range, using the CRV model as defined in Equation~\ref{eqn:mag} and then only optimizing the ROI, we obtain TS=23.8, while for the ICS model we find TS=12.4.
Note that the positions of other 4FGL sources are shown as white crosses. In cyan, we show the half-light radius of M54.\footnote{https://people.smp.uq.edu.au/HolgerBaumgardt/globular/}

\begin{figure*}
\centering
    \includegraphics[width=\textwidth]{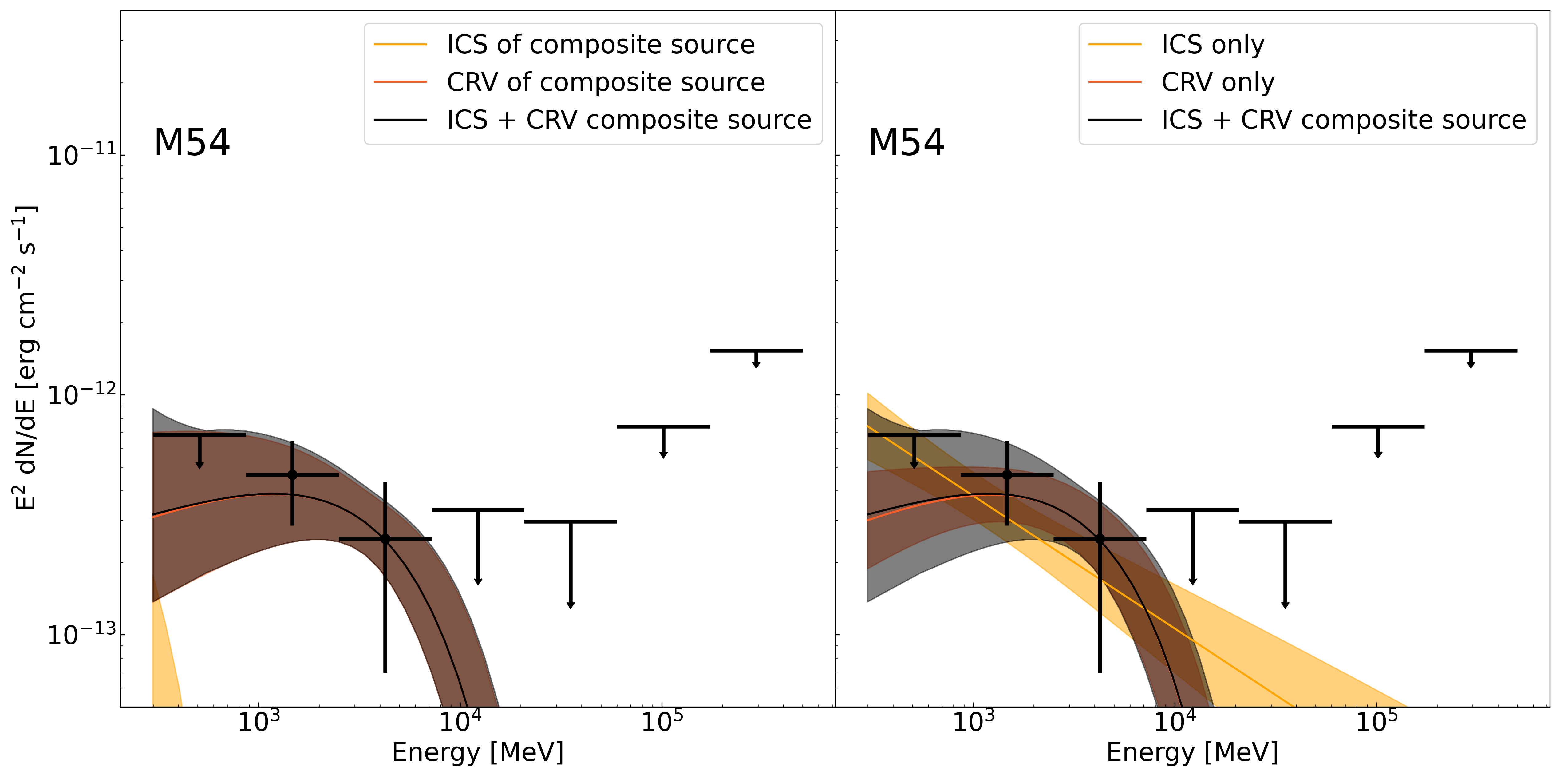}
    \caption{The spectral energy distribution (SED) of M54. The lines represent the best-fit models and the bands show the 1-sigma uncertainties on the model parameters. \textbf{Left:} The SED obtained in our two-component analysis, where the black curve represents the best fit total spectrum (see Equation~\ref{eqn:combo}), and the orange and yellow curves represent the components associated with curvature emission (CRV) and inverse Compton scattering (ICS). The ICS component is generally found to be negligible in this case, while the CRV component is well-defined. \textbf{Right:} the SED obtained for our one-component analysis (where the emission is considered to be either described by Equation~\ref{eqn:ics} or described by Equation~\ref{eqn:mag}). For comparison, we show again in this frame the best-fit two-component model in black.
    \label{fig:M54_sed}}
\end{figure*}

\begin{figure*}
\centering
    \includegraphics[width=\textwidth]{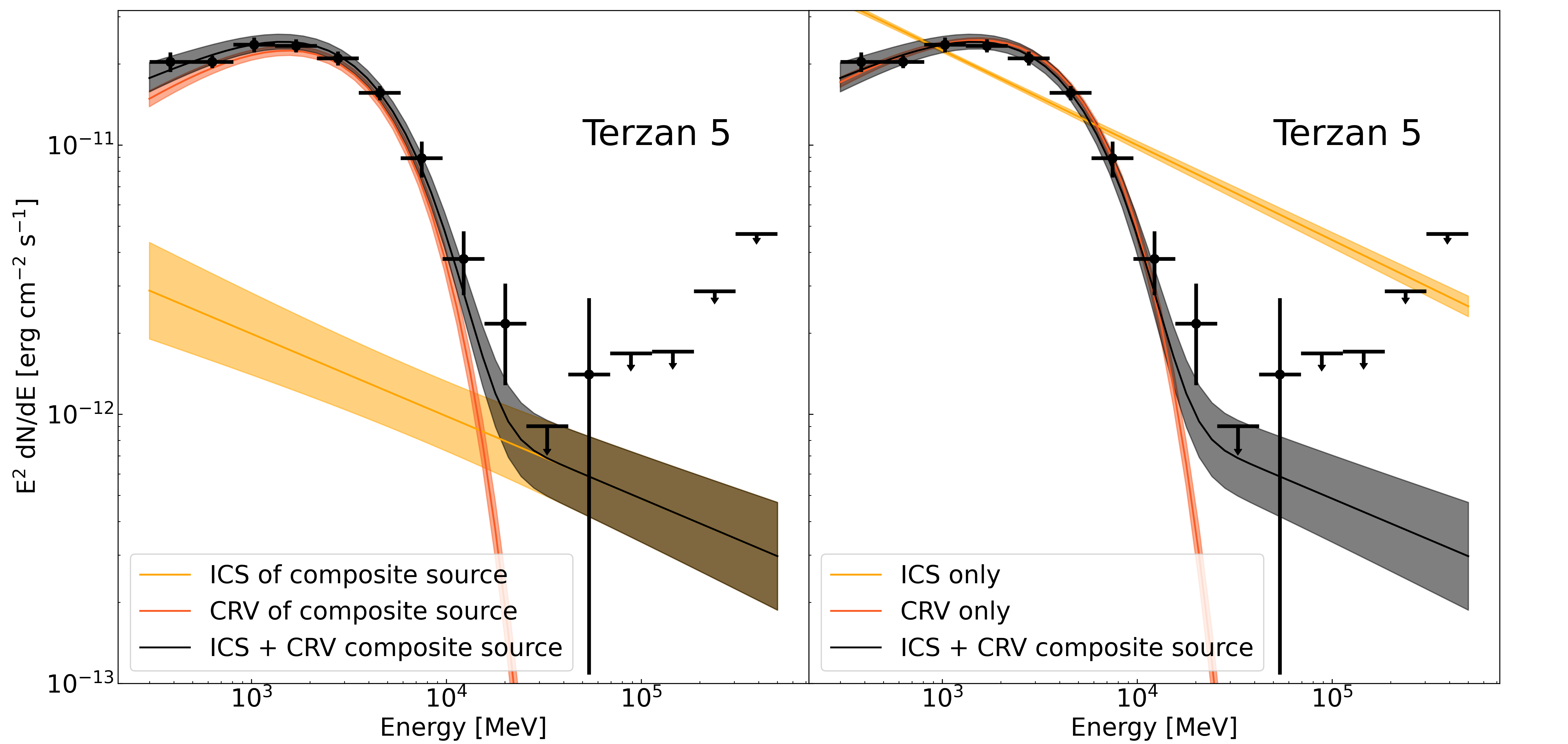}
    \caption{As in Figure~\ref{fig:M54_sed}, but for the globluar cluster Terzan 5. Unlike M54, this source has distinct contributions from both curvature emission (CRV) and inverse Compton scattering (ICS).}
    \label{fig:ter5_and_wcen_sed}
\end{figure*}   
\subsection{Point source spectral energy distributions}
\label{sec:SED} 

We show the results of our SED analysis of M54 in Figure~\ref{fig:M54_sed}~for the energy range of [300 MeV, 500 GeV]. Fitting with either a single power law source or a single CRV source yields similar results, with a TS of 18.9 and 17.50, respectively. While these TS values are slightly lower than the $> 100$ MeV analysis quoted in the previous section, this difference does not qualitatively change any of the subsequent interpretations. The spectral parameters we derive for each model are listed in Table~\ref{tab:spectralParams}. While the curved and power law models give statistically similar fits on their own, it is clear that there is no detected emission above $\sim$ 10 GeV and thus there is no ICS component in the two-component modeling.
\begin{table}
\centering
\resizebox{\columnwidth}{!}{
\begin{tabular}{cc|}
\hline
\multicolumn{1}{|c|}{\textbf{Model name}} & \textbf{Parameters} \\ \hline
\multicolumn{2}{|c|}{\textbf{M54}} \\ \hline
\multicolumn{1}{|c|}{\multirow{2}{*}{\textbf{ICS only}}} & $\gamma_1 = -2.55 \pm 0.21 $ \\
\multicolumn{1}{|c|}{} & $N_1$ = $[3.79 \pm 0.96] \times 10^{-7}$ cm$^{-2}$ s$^{-1}$ erg$^{-1}$ \\ \hline
\multicolumn{1}{|c|}{\multirow{3}{*}{\textbf{CRV only}}} & $\gamma_2 = -1.63 \pm 0.42$ \\
\multicolumn{1}{|c|}{} & $N_2 = [5.10 \pm 2.16] \times 10^{-7} $ cm$^{-2}$ s$^{-1}$ erg$^{-1}$ \\
\multicolumn{1}{|c|}{} & $E_\text{cut} = [3.38 \pm 2.15] \times 10^3$ MeV \\ \hline
\multicolumn{1}{|c|}{\multirow{5}{*}{\textbf{CRV + ICS of two-component source}}} & $\gamma_1 : $ Unconstrained \\
\multicolumn{1}{|c|}{} & $N_1 \leq$   \rm{ } $2.9 \times 10^{-8}$ cm$^{-2}$ s$^{-1}$ erg$^{-1}$ \\
\multicolumn{1}{|c|}{} & $\gamma_2 = -1.65 \pm 0.56$ \\
\multicolumn{1}{|c|}{} & $N_2 = [5.13 \pm 4.20] \times 10^{-7} $ cm$^{-2}$ s$^{-1}$ erg$^{-1}$ \\
\multicolumn{1}{|c|}{} & $E_\text{cut} = [3.44 \pm 2.33] \times 10^3$ MeV \\ \hline
\multicolumn{2}{|c|}{\textbf{Terzan 5}} \\ \hline
\multicolumn{1}{|c|}{\multirow{2}{*}{\textbf{ICS only}}} & $\gamma_1 = -2.35 \pm 0.02 $ \\
\multicolumn{1}{|c|}{} & $N_1$ = $[2.15 \pm 0.04] \times 10^{-5}$ cm$^{-2}$ s$^{-1}$ erg$^{-1}$ \\ \hline
\multicolumn{1}{|c|}{\multirow{3}{*}{\textbf{CRV only}}} & $\gamma_2 = -1.59 \pm 0.04$ \\
\multicolumn{1}{|c|}{} & $N_2 = [2.99 \pm 0.092] \times 10^{-5} $ cm$^{-2}$ s$^{-1}$ erg$^{-1}$ \\
\multicolumn{1}{|c|}{} & $E_\text{cut} = [3.63\pm 0.24] \times 10^3$ MeV \\ \hline
\multicolumn{1}{|c|}{\multirow{5}{*}{\textbf{CRV + ICS of two-component source}}} & $\gamma_1 = -2.35 \pm 0.09 $ \\
\multicolumn{1}{|c|}{} & $N_1$ = $[2.45 \pm 0.77] \times 10^{-6}$ cm$^{-2}$ s$^{-1}$ erg$^{-1}$ \\
\multicolumn{1}{|c|}{} & $\gamma_2 = -1.54 \pm 0.04$ \\
\multicolumn{1}{|c|}{} & $N_2 = [2.81 \pm 0.13] \times 10^{-5} $ cm$^{-2}$ s$^{-1}$ erg$^{-1}$ \\
\multicolumn{1}{|c|}{} & $E_\text{cut} = [3.24 \pm 0.23] \times 10^3$ MeV \\
\hline
\label{tab:spectralParams}
\end{tabular}
}
\caption{The best-fitting parameters and their uncertainties for the ICS, CRV, and CRV + ICS models. The corresponding SEDs for these fits are shown in Figures~\ref{fig:M54_sed}~and~\ref{fig:ter5_and_wcen_sed}.}
\end{table}

\par From this null detection of any ICS component, we can calculate an upper limit on the integrated ICS/high-energy flux. 
Integrating the ICS flux betweeen [300 MeV, 500 GeV], we find an upper limit for this component of $1 \times 10^{-13}$ erg cm$^{-2}$ s$^{-1}$. Comparing this to the flux observed in the CRV band, we obtain an upper limit of $L_{\text{ICS}} / L_{\text{CRV}} \leq 0.006 $.

\par
\begin{figure} 
\centering
    \includegraphics[width=\columnwidth]{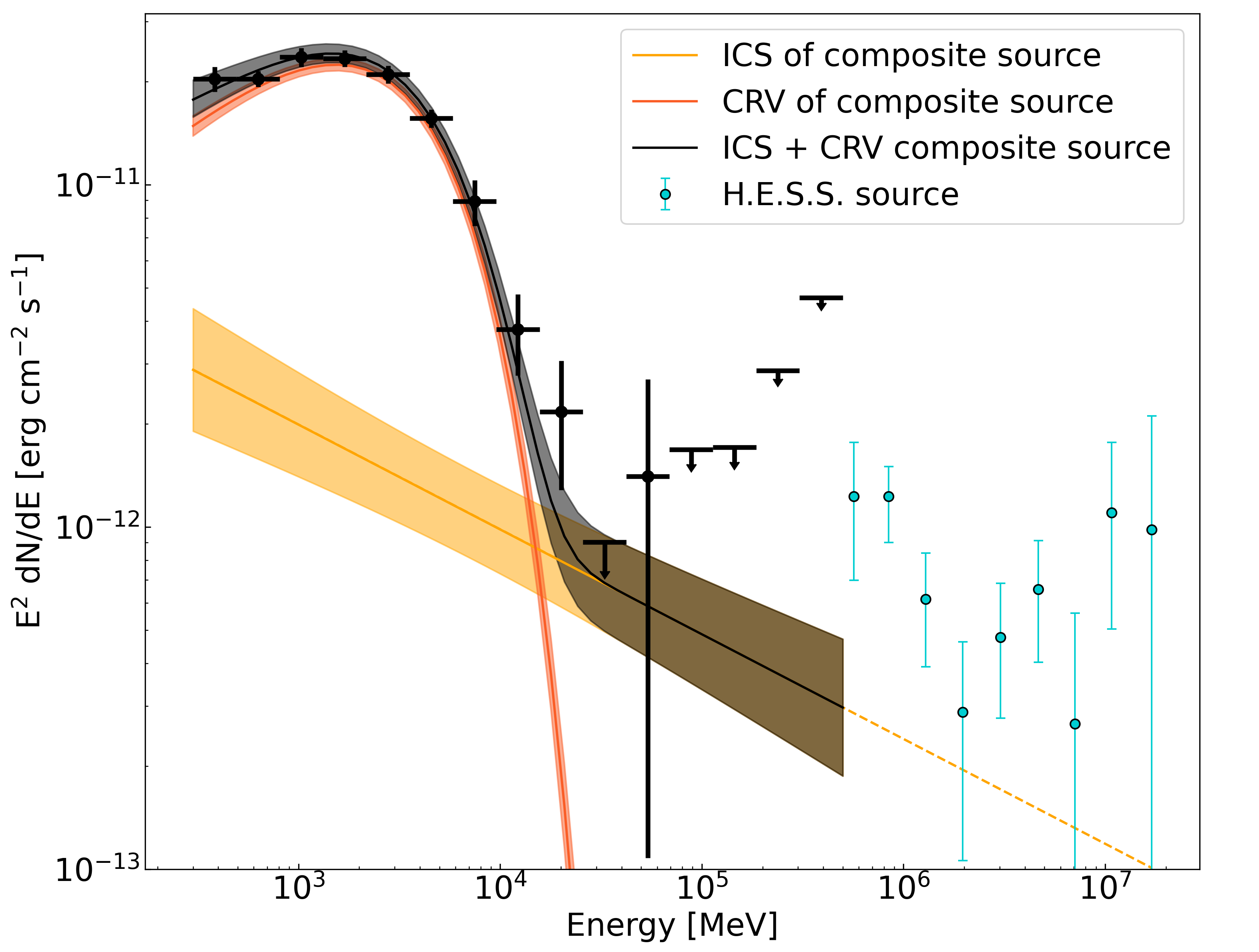}
    \caption{As in Figure~\ref{fig:ter5_and_wcen_sed}, but including measurements of Terzan 5 from the H.E.S.S. Collaboration~\citep{Ter5HESS}.}
    \label{fig:ter5_HESS_sed}
\end{figure}

 We can compare our results for Sgr/M54 to the well-studied case of Terzan 5 (as shown in Figure~\ref{fig:ter5_and_wcen_sed}).
For Terzan 5, our fit prefers the two-component model, featuring contributions from both CRV and ICS at a level given by $L_\text{ICS}$ / $L_\text{CRV} = 0.71 \pm 0.07$.   
We compare our results to the H.E.S.S. detection of Terzan 5 in Figure~\ref{fig:ter5_HESS_sed}.
These measurements from H.E.S.S. confirm the presence of a significant ICS component from this source, with a spectral index that is slightly harder than that favored by our analysis.

\subsection{Search for unassociated sources}
Using \texttt{gta.find\_sources}, we have identified 20 new sources with TS $>$ 9 within the 15$^\circ$ $\times$ 15$^\circ$ region centered on Sgr. 
From there, after checking the positions of the unassociated 4FGL sources and newly found sources, we find a total of 13 sources (3 new point sources and 10 catalog sources) that are within $1^\circ$ of a GC (not including the M54 source). We show the locations and the TS values of these sources in Table~\ref{tab:localization}.
\begin{figure*}
\centering
    \includegraphics[width=\textwidth]{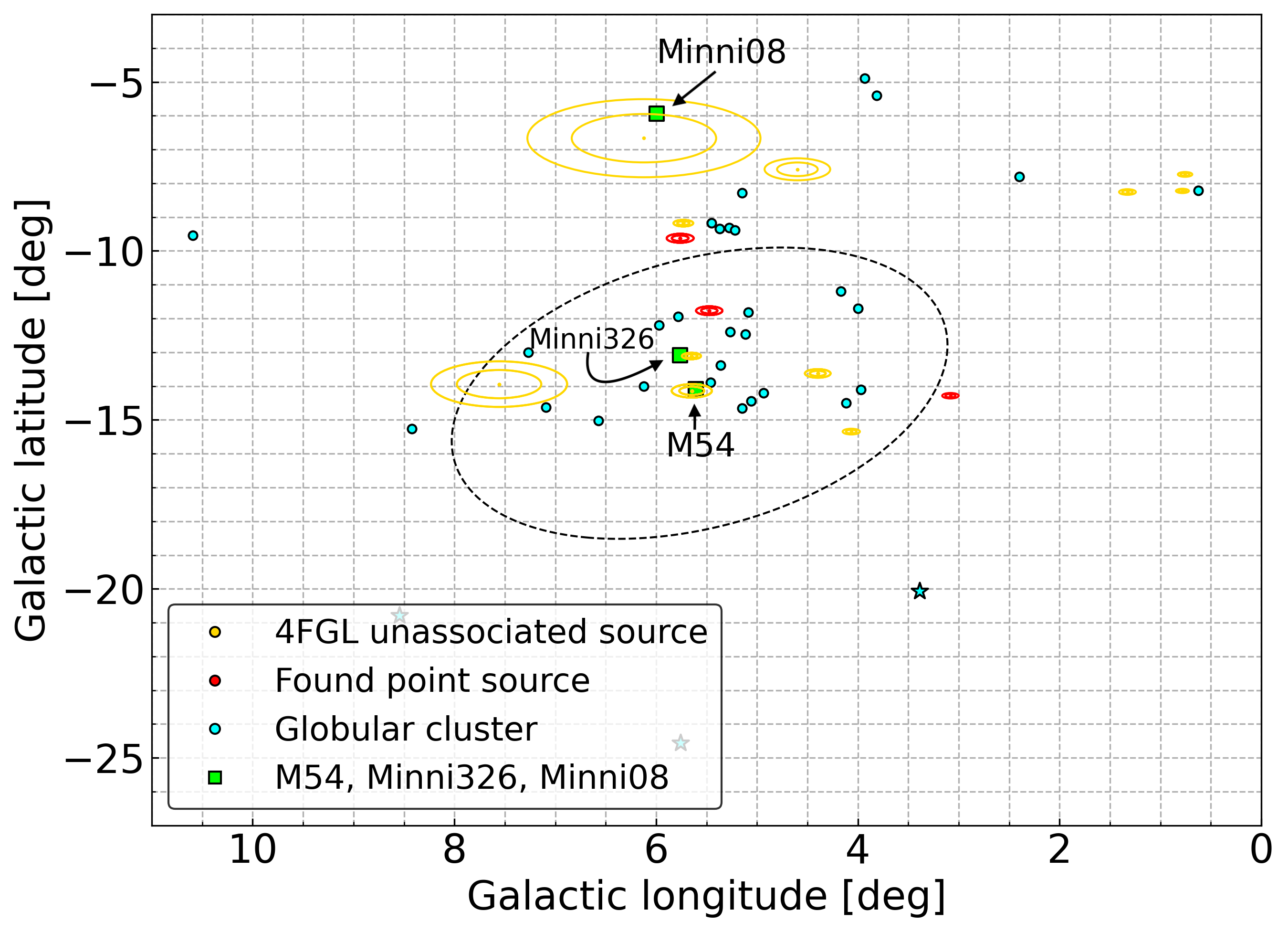}
    \caption{A map of the Sgr region. The blue points denote the location of Minni globular clusters associated with Sgr~\citep{Minniti2021_1, Minniti2021_2}, and the blue stars denote the previously known globular clusters within our ROI~\citep{2010AJ....140.1830G}. The red points represent the locations of the sources found using FermiPy's \texttt{gta.find\_sources} function, while yellow points are the locations of 4FGL-DR3 sources. The ovals represent the 1 and 2-sigma positional uncertainties on the locations as calculated from the \texttt{gta.localization} method. The green squares show the three globular clusters that have overlap with any of these sources, M54, Minni 326, and Minni 08. In black, we show the half-light radius of Sgr~\citep{Ferguson2020}. The TS values for all of the Fermi-LAT sources shown are given in Table~\ref{tab:localization}.
    \label{fig:minnis}}
\end{figure*}

\begin{table*}
\centering
\begin{tabular}{cccccc}
Source name & $l [^\circ]$ & $b [^\circ]$ & TS & Globular cluster name & \multicolumn{1}{l}{Distance to globular cluster {[}$^\circ${]}} \\ \hline
\multirow{7}{*}{\textbf{4FGL J1851.0-3003}} & \multirow{7}{*}{\textbf{5.65}} & \multirow{7}{*}{\textbf{-13.11}} & \multirow{7}{*}{\textbf{30.68}} & Minni148 & 0.39 \\
 &  &  &  & \textbf{Minni326} & \textbf{0.11} \\
 &  &  &  & Minni328 & 0.80 \\
 &  &  &  & Minni332 & 0.97 \\
 &  &  &  & Minni335 & 0.83 \\
 &  &  &  & Minni341 & 0.80 \\
 &  &  &  & M54 & 0.98 \\ \hline
\multirow{4}{*}{4FGL J1850.7-3216} & \multirow{4}{*}{4.39} & \multirow{4}{*}{-13.62} & \multirow{4}{*}{73.99} & Minni146 & 0.64 \\
 &  &  &  & Minni148 & 0.96 \\
 &  &  &  & Minni325 & 0.91 \\
 &  &  &  & Minni342 & 0.77 \\ \hline
4FGL J1857.8-3220 & 4.07 & -15.35 & 103.52 & Minni325 & 0.85 \\ \hline
\multirow{2}{*}{4FGL J1857.7-2830} & \multirow{2}{*}{7.56} & \multirow{2}{*}{-13.94} & \multirow{2}{*}{9.91} & Minni145 & 0.98 \\
 &  &  &  & Minni348 & 0.82 \\ \hline
\multirow{5}{*}{PS J1845.0-2939} & \multirow{5}{*}{5.47} & \multirow{5}{*}{-11.77} & \multirow{5}{*}{17.46} & Minni324 & 0.35 \\
 &  &  &  & Minni328 & 0.66 \\
 &  &  &  & Minni329 & 0.38 \\
 &  &  &  & Minni332 & 0.64 \\
 &  &  &  & Minni335 & 0.78 \\ \hline
PS J1851.3-3248 & 3.08 & -14.29 & 39.03 & Minni146 & 0.88 \\ \hline
\multirow{4}{*}{PS J1836.7-2829} & \multirow{4}{*}{5.76} & \multirow{4}{*}{-9.63} & \multirow{4}{*}{11.69} & Minni01 & 0.47 \\ 
 &  &  &  & Minni310 & 0.54 \\
 &  &  &  & Minni311 & 0.57 \\
 &  &  &  & Minni312 & 0.59 \\ \hline
\multirow{4}{*}{4FGL J1834.9-2819} & \multirow{4}{*}{5.73} & \multirow{4}{*}{-9.18} & \multirow{4}{*}{60.63} & Minni01 & 0.39 \\
 &  &  &  & Minni310 & 0.27 \\
 &  &  &  & Minni311 & 0.47 \\
 &  &  &  & Minni312 & 0.54 \\ \hline
4FGL J1826.2-2830 & 4.60 & -7.59 & 19.32 & Minni02 & 0.89 \\ \hline
4FGL J1822.0-3146 & 1.33 & -8.26 & 29.51 & Minni03 & 0.70 \\ \hline
\textbf{4FGL J1825.5-2647} & \textbf{6.12} & \textbf{-6.66} & \textbf{38.33} & \textbf{Minni08} & \textbf{0.73} \\ \hline
4FGL J1820.7-3217 & 0.78 & -8.22 & 38.33 & Minni03 & 0.16 \\ \hline
4FGL J1820.7-3217 & 0.77 & -7.74 & 121.12 & Minni03 & 0.50 \\ \hline
\end{tabular}
\caption{The results of our search for gamma-ray sources within the Sagittarius dwarf galaxy. "PS" denotes the point sources found using \texttt{gta.find\_sources}, while "4FGL" denotes unassociated catalog sources. We show here all PS and 4FGL sources found within $1^\circ$ of a globular cluster within the Sagittarius system. We use the locations of the Minni globular clusters as listed in~\citet{Minniti2021_1, Minniti2021_2}. The globular clusters which lie within the \texttt{FermiPy} localization uncertainties of the sources are shown in bold, and appear in Figure~\ref{fig:minnis} as green squares.}
\label{tab:localization}
\end{table*}

In Figure~\ref{fig:minnis}, we show a map of the GCs associated with Sgr, as well as nearby 4FGL catalog and other point sources.
After calculating the localization of each source within $1^\circ$ of a GC, we check if any GCs lie within the 68\% and 95\% containment regions of the sources' locations.
The containment regions for each source are shown as ovals of corresponding colors (yellow for catalog sources and red for new point sources). 
Besides M54, we find two sources that lie within the localizations of our unassociated sources: 4FGL J1851.3003 with Minni326 and 4FGL J1825.5-2647 with Minni08. Minni08's association with Sgr is inconclusive to date~\citep{Minniti2021_1}, and has no structural parameters determined thus far~\citep{Minniti2021_2}. For these reasons, we consider it unlikely that this GC is truly associated with a gamma-ray source.
In contrast, Minni326 is one of the brighter Minni GCs, with an estimated mass of $6.8 \times 10^3 \text{M}_\odot$. If this 4FGL source is associated with Minni326, this implies that Minni326 is simultaneously one of the furthest and least massive GCs to be detected in gamma-rays -- an unlikely combination. The TS of this 4FGL sources is $\sim 30$ for the power-law model, and $\sim 24$ for the case of the curved spectral model.

\par As mentioned before, there are no known pulsars in Sgr. While it is possible that there are other gamma-ray emitting sources within Sgr, this seems unlikely considering the masses of these GCs.
The least massive gamma-ray bright GC is GMS-01, with a mass of $3.5 \times 10^4 \text{M}_\odot$.
While one of Sgr's oldest known GCs, Terzan 8, possesses a slightly higher mass than this, the remainder of its $\sim 20$ GCs have masses at or below the mass of GMS-01.
Thus, while there are several spatial overlaps between unassociated gamma-ray sources and Sgr GCs, we do not suggest that they are associated.

\subsection{Extension tests on the M54 source}
We have checked for evidence of extension of the Sgr/M54 source, finding that the TS does not appreciably improve when using an extended template.


\begin{figure*} 
\centering
    \includegraphics[width=\textwidth]{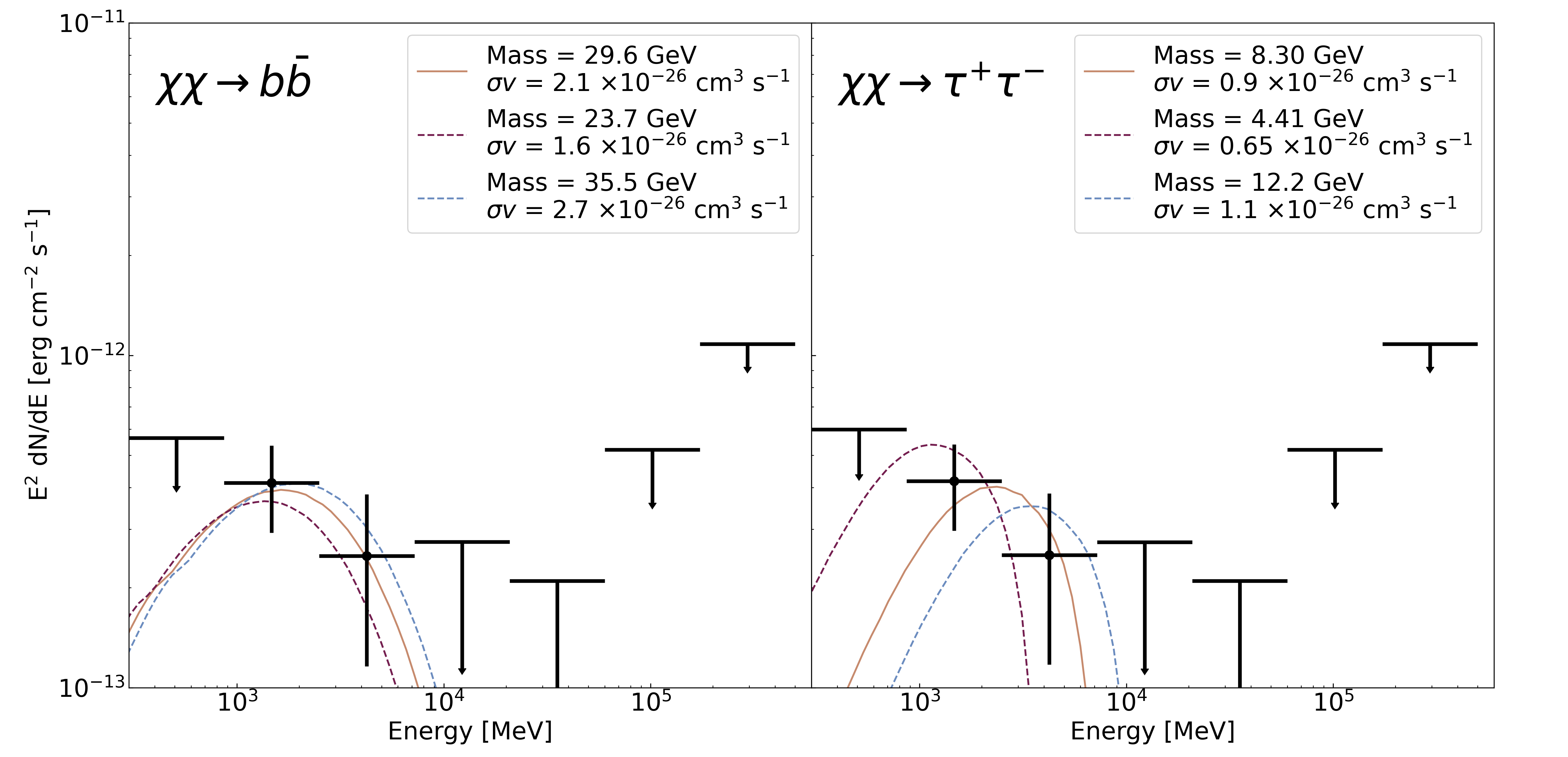}
     \caption{The SED of Sgr/M54 assuming that the gamma-ray emission is due to annihilating dark matter for the $b\bar{b}$ and $\tau^+ \tau^-$ channels. We show three lines for each channel: The solid tan line shows the annihilation spectrum of the best-fitting mass and annihilation cross section, $\sigma v$, based on the global fit to the source. The two dashed lines show the annihilation spectrum of the lower and upper edges of the 1-sigma uncertainty of the fits. We have adopted a J-factor of $10^{19.6}$ GeV$^{2}$ cm$^{-5}$.}
    \label{fig:dmM54}
\end{figure*}
\begin{figure*}
\centering
    \includegraphics[width=\textwidth]{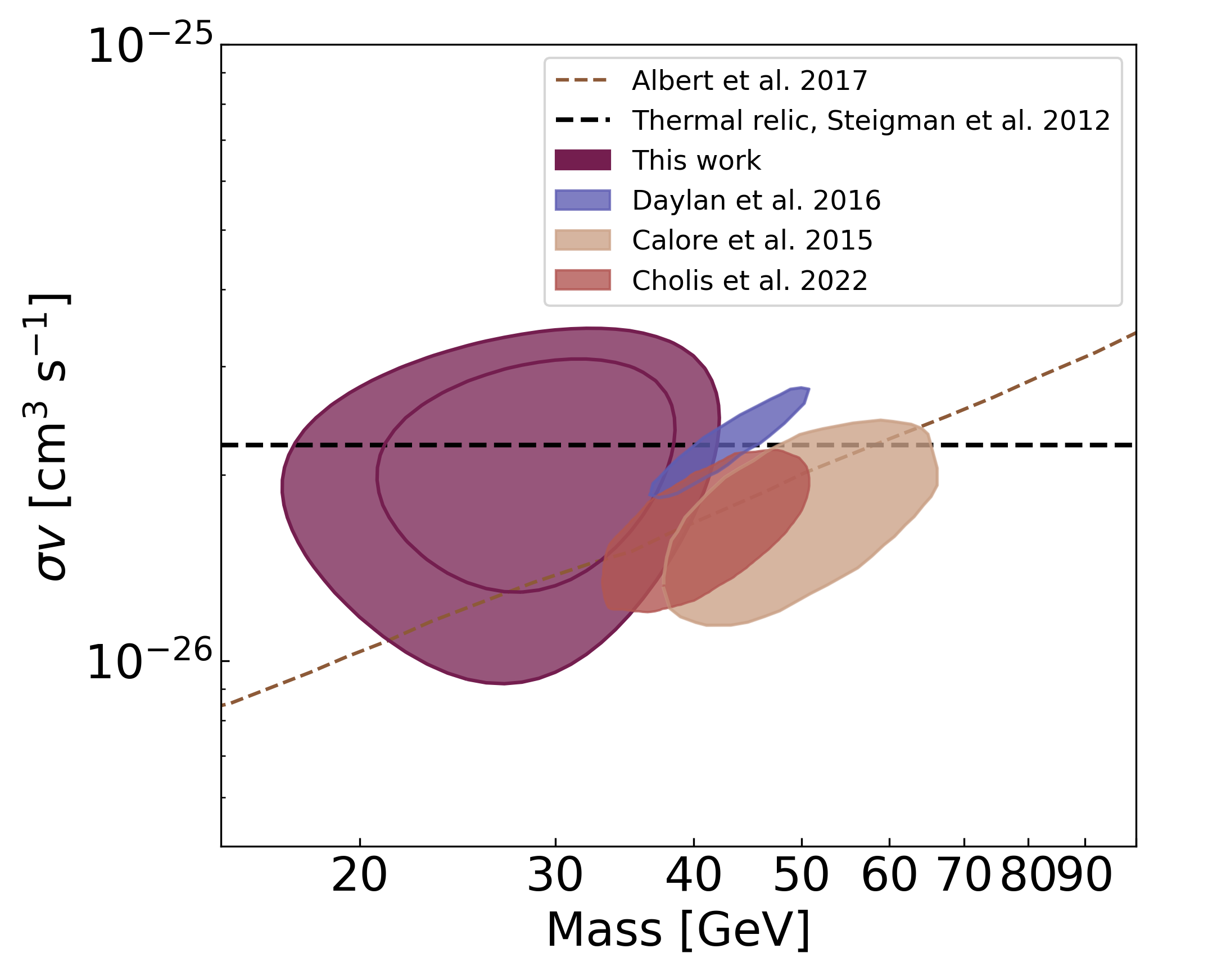}
    \caption{The regions of dark matter parameter space which provide a good fit to the gamma-ray flux observed from the core of the Sgr dwarf galaxy (purple), under the assumption that all of this emission originates from dark matter annihilation, and adopting a J-factor of $10^{19.6}$ GeV$^{2}$ cm$^{-5}$. For the contours corresponding to the results of this work, the dark lines represent the 68\% and 95\% containment regions. The black dashed line is the annihilation cross section that is predicted for a thermal relic~\citep{2012PhRvD..86b3506S}. The brown dashed line denotes the upper limit on the dark matter annihilation cross section from the null results of searches for gamma-ray emission from Milky Way dwarf galaxies~\citep{2017ApJ...834..110A}. The remaining contours show 2-sigma fits to the Galactic Center Gamma-Ray Excess~\citep{2016PDU....12....1D,2015JCAP...03..038C,Cholis2022}. All results shown in this figure are for the case of dark matter annihilations to $b\bar{b}$.}
    \label{fig:dmContours}
\end{figure*}

\section{Dark matter interpretation} 
\label{sec:dm}
\par Up to this point, we have assumed that the point source residing at the center of Sgr is associated with the GC, M54. However, it is prudent to also consider a scenario in which the gamma-ray emission from this dwarf is not from M54, but is instead from DM annihilating within the core of this dSph's DM halo.
The flux of gamma-rays from annihilating DM is given by,
    \begin{equation}
            \frac{d\Phi}{dE} = \
            \frac{1}{8\pi}
            \frac{<\sigma v>}{m_\chi^2}
            \frac{dN}{dE}
            \times J,
        \label{eqn:dm}
    \end{equation}
where $m_\chi$ is the mass of the DM particle, $\langle \sigma v \rangle$ is the velocity-weighted annihilation cross section, and $dN/dE$ is the flux density per annihilation, which depends on the DM's mass and annihilation channel(s).  $J$ is what is typically referred to as the ``astrophysical piece'' of the above equation because it depends on the density and morphology of the DM halo, which can be derived from kinematic measurements. The J-factor is given by:
    \begin{equation}
            J(\theta_\text{max}) = \int \int \rho^2_\text{DM}(r) \text{d}\ell \text{d}\Omega,
        \label{eqn:jfactor}
    \end{equation}
where $\ell$ is a line-of-sight through the halo and $\Omega$ is the solid angle with a radius, $\theta_\text{max}$. 
We refer the reader to~\citet{2019MNRAS.482.3480P} for a review of the methodology that we use for calculating J-factors from kinematic data. 

\par We measure the two components of the tangential velocity dispersion and the radial velocity dispersion of Sgr using data from Gaia EDR3 \citep{Gaia_Brown_2021A&A...649A...1G} and APOGEE DR16 \citep{SDSS_DR16_2020ApJS..249....3A}. 
We identify 778 Sgr members in the innermost 125\arcmin~based on stars consistent with the systemic line-of-sight velocity,  proper motion, and parallax of Sgr \citep{2020MNRAS.497.4162V}.
Adopting an NFW DM profile,
\begin{equation}
    \rho(r) = \frac{\rho_s}{\frac{r}{r_s} \left( 1+\frac{r}{r_s} \right)^2},
\end{equation}
we measure the posterior probability density functions for the scale density, $\rho_s$, and then convert this into a J-factor. We fix the scale radius to be $r_s = 1$ kpc, which corresponds to the approximate half-light radius of Sgr. From this method, we determine the integrated J-factor within the half-light radius ($\sim$ 2 degrees of the Sgr core) to be
$\log_{10} [{\rm J} ({\rm GeV}^{2} {\rm cm}^{-5})] = 19.6 \pm 0.2$. 
In calculating this quantity, we have assumed that the core region of Sgr is in dynamical equilibrium. 
If this assumption is not valid, there could be large systematic uncertainties on the J-factor, beyond those we have quoted above. We address this issue in more detail in the discussion below.

\par For a halo with a scale radius of $r_s \sim 1 \, {\rm kpc}$, DM annihilation in Sgr should be expected to produce a gamma-ray signal that is detectably extended. Before considering that case, however, we first present our results for the case of point-like emission, as shown in Figure~\ref{fig:dmM54} for the $b \bar{b}$ and $\tau^+ \tau^-$ annihilation channels.
For the $b \bar{b}$ channel, we find that this spectrum is best fit by a DM particle with a mass of $m_\chi= 29.6 \pm 5.8$ GeV and an annihilation cross section of $\sigma v = (2.1 \pm 0.59) \times 10^{-26} \text{cm}^3/$s. 
For the $\tau^+ \tau^-$ channel, we find that the fit prefers $m_X=8.3 \pm 3.8$ GeV and $\sigma v = (0.90 \pm 0.25) \times 10^{-26}~\text{cm}^3/{\rm s}$. In these cases, we obtain TS=16.5 ($b\bar{b}$) and TS=16.2 ($\tau^+ \tau^-$). As was the case for the GC models, if we extend our fitting down to 100 MeV we recover a higher TS of 22.9 and 18.3, respectively. In Figure~\ref{fig:dmContours}, we show the regions of the dark matter parameter space that are favored for DM annihliating to $b\bar{b}$. This region, shown in purple, was derived from the full covariance matrix in the space of $m_\chi$ and $\langle \sigma v \rangle$. 
Other results, including the regions favored by the observed properties of the Galactic Center Gamma-Ray Excess, are shown for comparison~\citep{2017ApJ...834..110A,2016PDU....12....1D,2015JCAP...03..038C,Cholis2022}.

To consider the possibility of detecting annihilation products from an extended DM halo, we construct spatial templates using an NFW density profile to describe the emission, and refer the reader to~\citet{2015JCAP...09..016H} for more details regarding the template construction. 
We define our templates out to a radius of $6^{\circ}$ from the center or Sgr, and cut the extended emission off beyond a radius of $2$ kpc. For very small values of $r_s$, we recover the results obtained for the point-like template, as described above. For larger values of $r_s$, however, we obtain smaller values for the TS. In particular, for $r_s=0.01$, 0.1, and 1 kpc, we find TS values of 12.3, 6.5, and 6.5, respectively.\footnote{Recall that we adopted $r_s =$ 1 kpc in deriving the J-factor from stellar kinematics as described above.} The fit thus prefers templates that are not significantly extended, somewhat disfavoring DM interpretations of this signal.

\section{Discussion and conclusion} 
\label{sec:final}
In this paper, we have analyzed the core of the Sagittarius dwarf spheroidal galaxy using data from Fermi-LAT. We have confirmed the existence of point-like emission from this region, which is associated with the globular cluster, M54, in the 4FGL-DR3 catalog. We find no evidence for emission from this source at energies  $\gtrsim 10$ GeV. If this emission originates from MSPs, this result suggests that it is produced largely at the pulsars' magnetosphere, and not through Inverse Compton scattering. We also consider a dark matter interpretation of this data, and derive values for the particle mass and annihilation cross section which provide a good fit to this signal. Testing both the $b\bar{b}$ and $\tau^+ \tau^-$ channels, we find best-fitting masses and cross sections which are consistent with the Galactic Center Gamma-Ray Excess, and with previous constraints from observations of other dwarf galaxies.

\subsection{Globular cluster interpretation} 
\label{sec:gcdiscuss}
\begin{figure*}
\centering
    \includegraphics[width=\textwidth]{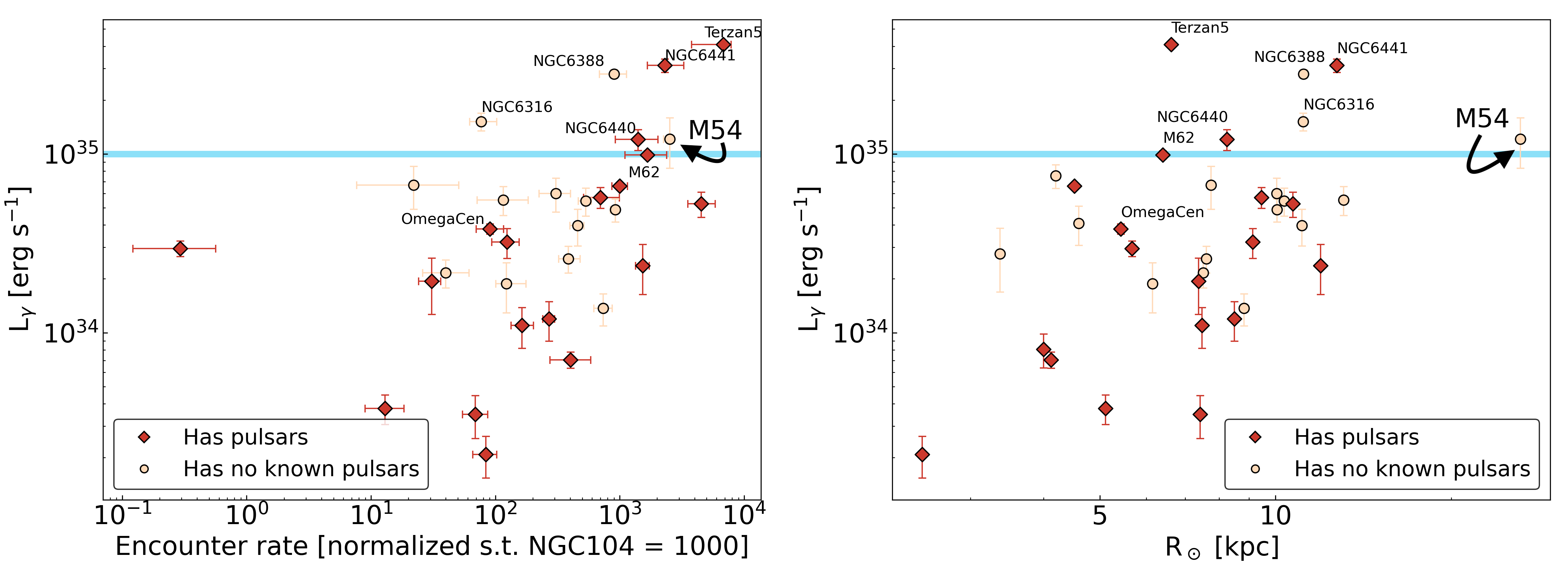}
    \caption{A comparison of M54 to other gamma-ray bright globular clusters. We denote GCs with (without) known pulsars by a dark red diamond (off-white circle). The left-hand panel shows the gamma-ray luminosity versus stellar encounter rate (see text for details), normalized such that NGC104 has an encounter rate of 1000. The right-hand panel shows the distance to the globular clusters from the Sun.}
    \label{fig:fermiGCs}
\end{figure*}

\begin{figure*} 
\centering
    \includegraphics[width=\textwidth]{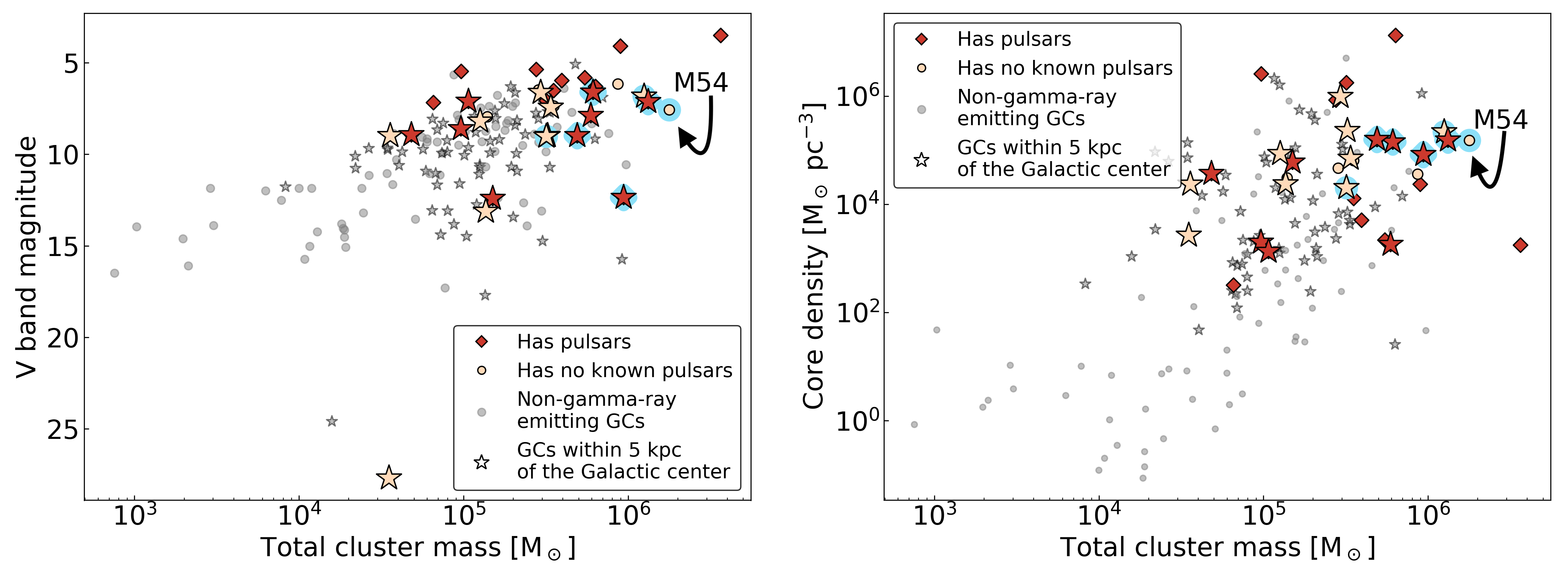}
    \caption{A comparison of all Milky Way GC masses and their V-band magnitudes (left-hand panel) as well as their core densities (right-hand panel). All values shown are from the Baumgardt Globular Cluster Database. Gamma-ray bright GCs are shown as in Figure~\ref{fig:fermiGCs} and our sub-population of high gamma-ray luminosity GCs are denoted by a blue outline around the marker. We also show which GCs are within 5 kpc of the Galactic center with the star symbol. Grey points on the figure denote GCs which have not been detected in gamma-rays.}
    \label{fig:fermiGCs_mass}
\end{figure*}

In Figures~\ref{fig:fermiGCs}~and~\ref{fig:fermiGCs_mass}, we compare the characteristics of M54 to those of 35 other gamma-ray bright GCs, highlighting those with a gamma-ray luminosity of  L$_\gamma \ge 10^{35}$ erg s$^{-1}$ (Terzan 5, NGC 6388, NGC 6316, NGC 6440, M62, and NGC 6441), and indicating which are known to contain pulsars.\footnote{http://www2.naic.edu/~pfreire/GCpsr.html} In Figure~\ref{fig:fermiGCs}, we plot the gamma-ray luminosities of these GCs against  their stellar encounter rate (as reported by~\citet{Bahramian2013}), and their distance from the Solar System. 
The encounter rate depends on the core density and core radius of the GC such that a more compact GC will have a higher encounter rate and thus a higher number of binary interactions.
This quantity has been shown to correlate with the predicted (observed) number of millisecond pulsars (X-ray binaries) within the cluster, e.g.~\citet{2003A&A...403L..11G, 2019MNRAS.486..851D}.
In~\citet{Bahramian2013}, the stellar encounter rates are estimated from the observed surface brightness profiles of the systems.
Upon deprojection of the surface brightness profiles, a luminosity density function can be derived. 
From this, the encounter rate of a GC is defined as,
\begin{equation}
\Gamma_e = \int \frac{\rho^2(r)}{\sigma_c} dV, 
\end{equation}
where $\rho$ is the stellar density profile of the cluster and $\sigma_c$ is the velocity dispersion at the core radius.
From this figure, we see that the observed gamma-ray luminosity of M54 is unsurprising in light of its large stellar encounter rate. This fact favors the hypothesis that this source's gamma-ray emission originates from MSPs. In the right panel of Figure~\ref{fig:fermiGCs}, we see that if this source is indeed associated with M54, then this is the most distant GC to have been detected by Fermi-LAT.
We also note that M54 has one of the highest X-ray fluxes of all globular clusters~\citep{2006Ramsay_M54}, suggesting a high number of X-ray binaries, the progenitor systems to MSPs.

\par One might expect that the most massive GCs, such as~\wcen, would have the highest gamma-ray luminosities. This, however, is not necessarily the case. 
From Figure~\ref{fig:fermiGCs_mass}, we see that while all of the gamma-ray bright GCs have high masses, densities, and magnitudes, there are several other GCs with similar properties that have not been detected by Fermi-LAT. 
In fact, there are other GCs with stellar masses as large as $\sim 10^6 \text{M}_\odot$ which are not gamma-ray bright, including NGC 2419, Liller 1 (see, however,~\citet{2011Tam}), 
NGC 5824, and NGC 6273.

\par Our analysis has not identified any evidence of emission above 10 GeV from M54. This could be considered surprising in light of the fact that TeV halos appear to be a universal feature of young and middle-aged pulsars~\citep{Hooper:2017gtd,Linden:2017vvb,HAWC:2021dtl,HAWC:2019tcx}, and perhaps millisecond pulsars as well~\citep{Hooper:2021kyp}. As previously discussed, Terzan 5 is the only GC that has been robustly detected at TeV-scale energies. The morphology of this emission is not entirely understood, however, as it extends beyond the tidal radius of this source and beyond the point spread function of H.E.S.S. Moreover, this TeV emission is offset from the center of the cluster by $\sim 4$ arcminutes. With future telescopes, such as the Cherenkov Telescope Array, it may be possible to detect the extended TeV-scale emission from the MSP populations within GCs~\citep{2019PhRvD.100d3016S}.  

\subsection{Dark matter interpretation}
\label{sec:dmdiscussion}

In Figure~\ref{fig:dmContours}, we show the DM parameter space that could explain the spectrum and intensity of the gamma-ray emission observed from the direction of M54. These results are consistent with the measured characteristics of the Galactic Center Gamma-Ray Excess, and with all existing constraints. 

There are several systematic uncertainties that one should keep in mind when considering these results. First, we have adopted a J-factor of $10^{19.6}$ GeV$^{2}$cm$^{-5}$ for Sgr. To calculate such a J-factor requires a Jeans analysis of the stellar kinematics, which relies on the assumption that the system in question is in dynamical equilibrium. This is not obviously true in the case of Sgr. In~\citet{2022arXiv220612121W}, the authors identify Sgr-like systems in the \texttt{AURIGA} simulations and test the accuracy of Jeans modeling to extract the actual mass of the dSph. The authors found that the masses of Sgr-like systems were systematically underestimated if the Jeans analysis was performed within the inner 200-300 pc of the dSph. Extrapolating this to a larger radius would suggest an overestimation of the J-factor for the analysis performed in this study. 
The value of the scale radius of Sgr's DM halo is also an important source of uncertainty. In this analysis, we have adopted a value of $r_s = 1 \, {\rm kpc}$, which matches the half-light radius of this system. However, the Jeans analysis still allows for the possibility that $r_s$ could be larger. If this is the case, fits to the kinematic data would prefer lower halo densities and thus smaller values of Sgr's J-factor.

Another interesting feature of M54 is its large central velocity dispersion~\citep{2009ApJ...699L.169I}, which could indicate the presence of a significant fraction of dark mass near the core of this GC. These high velocity dispersions were originally attributed to a possible intermediate mass black hole candidate; see however, the most recent analysis of~\citet{2011AJ....142..113W}.
Other possibilities include the dark mass being a population of stellar remnants which have sunk to the core of the system due to mass segregation~\citep{2020ApJS..247...48K}, or the central cusp of Sgr's DM halo~\citep{2022ApJ...935...14C}. \citet{2022ApJ...935...14C} have argued that tidal effects may have greatly disturbed Sgr's DM halo, leaving only the innermost $\sim$300 pc relatively unperturbed.
This could explain why the gamma-ray emission from this source is approximately point-like, showing no signs of spatial extension.

\par Dedicated pulsar searches in the radio band, as well as searches for gamma-ray pulsations, could shed significant light on the nature of the Sgr/M54 system. Recently,~\citet{2022RAA....22k5013Y} performed a study of a number of gamma-ray bright GCs, including M54.
The authors searched for pulsations in Fermi-LAT data and found no evidence for time-dependent variations in M54's flux.
As more data is acquired by Fermi-LAT, it may be possible to detect pulsations associated with the M54 source. 
Dedicated radio searches, such as with the Square Kilometer Array, may also find pulsars in either the main body of Sgr, or within M54~\citep{2015aska.confE..40K}.

\section*{Acknowledgements}
AJE and LES acknowledge support from DOE Grant de-sc0010813. This work was supported by the Texas A\&M University System National Laboratories Office and Los Alamos National Laboratory, and NASA grant number 80NSSC22K1577. AJE acknowledges support from the Texas Space Grant Consortium. 
AJE thanks all organizers, instructors, and students of the 2022 Fermi Summer School for useful discussions, and especially Elizabeth Hays, Matthew Kerr, and Jamie Holder. DH is supported by the Fermi Research Alliance, LLC under Contract No.~DE-AC02-07CH11359 with the U.S. Department of Energy, Office of Science, Office of High Energy Physics. TL is supported by the European Research
Council under grant 742104, the Swedish Research Council under contract 2019-05135 and the
Swedish National Space Agency under contract 117/19.
ABP is supported by NSF grant AST-1813881.
AA and JPH are supported by the US Department of Energy Office of High-Energy Physics and the Laboratory Directed Research and Development (LDRD) program of Los Alamos National Laboratory.
\par This work was completed by the use of the \textsc{Python} programming language as well as the following software packages: \textsc{astropy} ~\citep{astropy:2018}, \textsc{pandas}~\citep{reback2020pandas}, \textsc{numpy}~\citep{         harris2020array}, \textsc{scipy}~\citep{2020SciPy-NMeth}, \textsc{matplotlib}~\citep{Hunter:2007}, and \textsc{spyder}~\citep{raybaut2009spyder}.

\bibliographystyle{mnras}
\bibliography{main} 

\bsp	
\label{lastpage}
\end{document}